\documentclass[%
aps,
reprint,
amsmath,amssymb,
pra,
]{revtex4-2}

\usepackage{graphicx}% Include figure files
\usepackage[export]{adjustbox}
\usepackage{cleveref}
\Crefname{figure}{Fig.}{Figs.}
\usepackage{siunitx}

\begin{document}

%\preprint{APS/123-QED}

\title{Inverse solving the Schrödinger equation for precision alignment of a microcavity}

\author{Charlie Mattschas, Marius Puplauskis, Chris Toebes, Violetta Sharoglazova, Jan Klaers}
\affiliation{%
 Adaptive Quantum Optics (AQO),\\
 MESA+ Institute for Nanotechnology, University of Twente, PO Box 217, 7500 AE Enschede, Netherlands % \textbackslash\textbackslash
}%


%\affiliation{
%\affiliation{
%\affiliation{%

\date{\today}% It is always \today, today,

%\justifying

%%%%%%%%%%%%%%%%%%%%
%%%   ABSTRACT   %%%
%%%%%%%%%%%%%%%%%%%%

\begin{abstract}

In paraxial approximation, the electromagnetic eigenmodes inside an optical microresonator can be derived from a Schrödinger-type eigenvalue problem. In this framework, tilting the cavity mirrors effectively introduces a linear potential to the system. In our work, we apply solution strategies for inverse problems to precisely determine and control the relative orientation of two mirrors forming an optical microcavity. Our approach employs the inversion of the Schrödinger equation to reconstruct the effective potential landscape, and thus mirror tilts, from observed mode patterns. We investigate regularization techniques to address the ill-posed nature of inverse problems and to improve the stability of solutions. Our method consistently achieves an angle resolution of order 100 nanoradians per measurement.

%\begin{description}
%\end{description}
\end{abstract}

\maketitle

%\tableofcontents

%%%%%%%%%%%%%%%%%%%%%%%%
%%%   INTRODUCTION   %%%
%%%%%%%%%%%%%%%%%%%%%%%%

\section{Introduction}  \label{sec:intro}

Inverse problems in physics represent a class of problems where the goal is to determine the underlying parameters of a system based on observed effects
\cite{Franceschetti1999,Sanchez2018,Pakrouski2020,Breton2024,Liu2024}. 
They are crucial in a wide range of physical applications, from imaging techniques \cite{Lan2024,Virupaksha2024,Zhou2024,Xu2024} to the study of quantum systems \cite{Inui2023,Weymuth2014,Yu2021}.
At their core, inverse problems involve formulating a mathematical model that relates the unknowns of a system
to observable data and typically involve a combination of analytical and numerical methods \cite{badia2024,Alpar2024,Zunger2018}. 
This reversal of the usual computational direction presents unique challenges, as inverse problems are often ill-posed, meaning they do not necessarily guarantee the existence, uniqueness, or stability of a solution. 
Small errors in the observed data can lead to significant deviations in the inferred causes, making the solution process inherently challenging. 
To address these issues, regularization techniques are employed, which involve introducing additional information or constraints to the problem. This helps in stabilizing the solution and making it more robust to data inaccuracies. 

In our work, we apply inverse problem solution strategies for determining the relative orientation of two cavity mirrors. Optical microcavities have played a pivotal role in a wide range of applications, including lasers, optical sensors, and quantum devices, for several decades \cite{vahala2003, antoniadis2023, ossiander2023, Armani2003, Barland2002, Santori2002}. Particularly for open microcavities, aligning both the cavity length and the relative orientation of the two mirrors can present a significant challenge \cite{Li2019,Niu2020,Kaiser2018,zhou2023,pallmann2023}. Useful methods for tackling this problem include deflection measurements (autocollimation) and interferometry. The latter is widely used for aligning purposes, high-precision surface profiling and related tasks \cite{Li2021, Duan2021, Pavlicek2020, Zhang2020, Lee1990, Kino1990, Dresel1992, Caber1993}. 
However, in certain experimental situations, such techniques might not be available or the best solution for quantitative assessment of cavity alignment. 
Here, we approach the alignment of a microcavity as an inverse problem. Through optical excitation with a laser, we create and observe spatially confined mode patterns within the microcavity, which are then analyzed to determine the relative angles of the two mirrors. The underlying model connecting the observations with system parameters is based on a Schrödinger equation derived from a paraxial treatment of light propagation in optical resonators. We introduce and compare several methods of regularization to assess their efficacy. We expect that a main application area of our method will be research on two-dimensional photonic and polaritonic gases in microresonators \cite{Kalinin2022,Inagaki2016,Berloff2017,Alyatkin2020,Stroev2020,Stroev2021}.

%%%%%%%%%%%%%%%%%%
%%%   THEORY   %%%
%%%%%%%%%%%%%%%%%%

\section{Method}    \label{sec:method}

\begin{figure*}[t] 
    \includegraphics[width=0.9\textwidth]{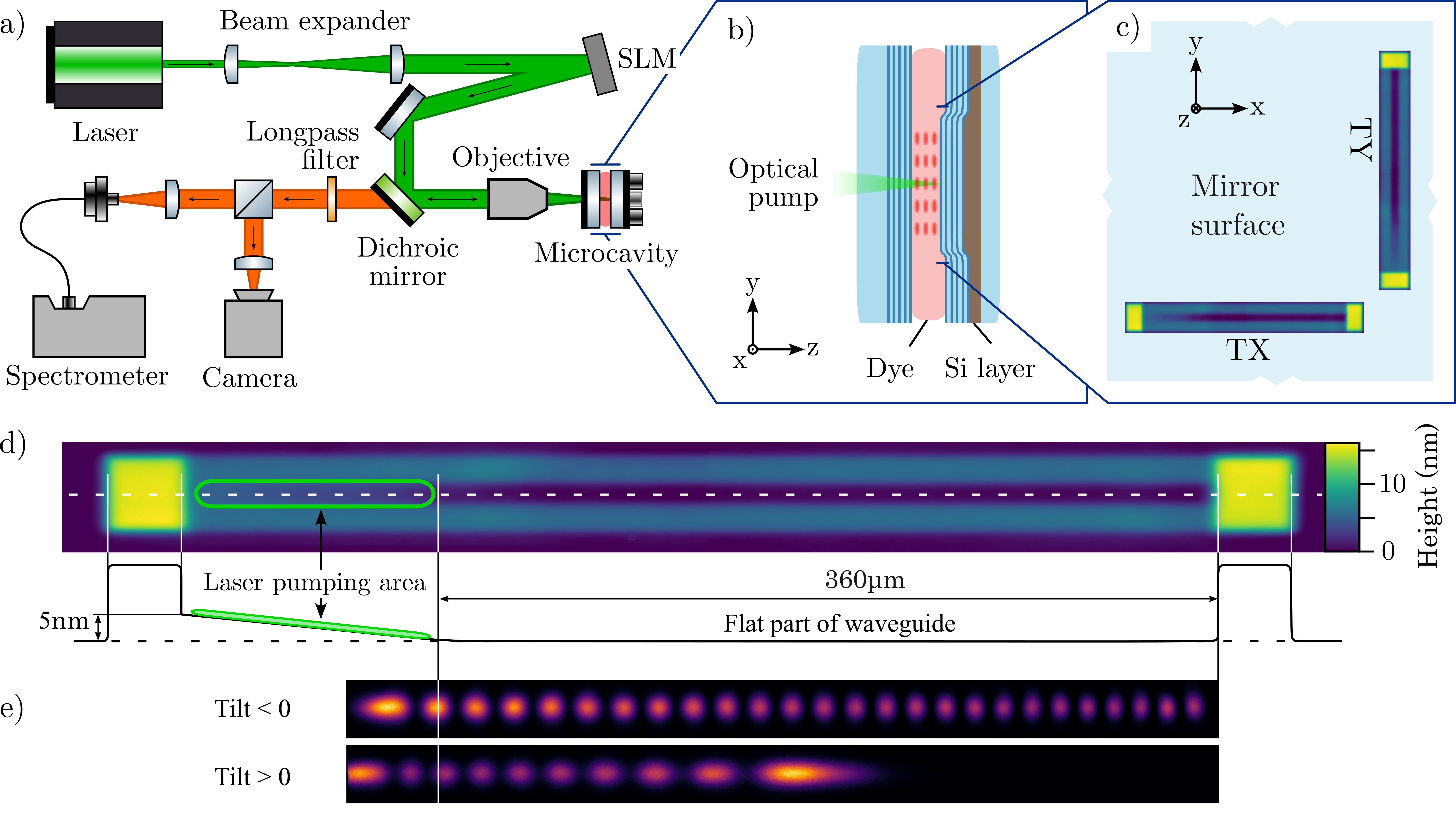}
    \caption{(a) Schematic of the experimental setup. The experiments are conducted in a microcavity consisting of two optical mirrors with a typical separation of approximately $\SI{10}{\micro m}$ and an optically active medium (rhodamine 101 in ethylene glycol). A non-resonant laser beam at a wavelength of $\SI{532}{\nano m}$ is used to generate light emission inside the cavity. The pump beam is modulated using a spatial light modulator (SLM) to control the position of the pump spot within the cavity plane. The transmitted signal at a wavelength of approximately $\SI{650}{\nano m}$ is split with a beam splitter and measured using a camera and a spectrometer. (b) Zoomed-in schematic of the microcavity. One of the mirrors includes an additional silicon (Si) layer between the dielectric stack and the superpolished substrate. This silicon layer is utilized to create arbitrary confining structures on the mirror surface using a direct laser-writing technique \cite{vretenar2023}. (c) Top-view schematic of the nanostructured mirror surface (the $xy-$plane of the mirror). The tilting angles in the $x-$ and $y-$directions are determined by analyzing the light emission in two orthogonal waveguide structures, TX and TY. (d) Height scan of a waveguide structure with a cross-section through the structure along the white, dashed line, shown below. This depicts a closed, linear waveguide with a ramp on the left side. We scan the position of the pump spot along this ramp and analyze the resulting intensity distribution from the flat part of the waveguide for the tilt determination. (e) Camera images showing mode patterns for negative and positive tilts of the mirror.}
    \label{fig:setup}
\end{figure*}

In a paraxial approximation, the photonic eigenmodes inside a microresonator can be derived from a Schr\"odinger-type eigenvalue problem \cite{Gloge1969,Vretenar2021}
\begin{align} \label{eq:potential}
    E\Psi(x,y) = -\frac{\hbar^2}{2m}\nabla^2\Psi(x,y) +V(x,y)\Psi(x,y),
\end{align}
where $\Psi(x,y)$ is the (scalar) wavefunction, $m$ is an effective photon mass, $V(x,y)$ is a potential energy term, and $E$ is the eigenenergy.
In such a framework, the potential energy term $V(x,y)$ relates to the local mirror separation in the cavity $D_0+\Delta d(x,y)$ with $\Delta d(x,y)\ll D_0$ in the following way:
\begin{align} \label{eq:potential}
    V(x,y) = -\frac{mc^2}{n^2} \frac{\Delta d(x,y)}{D_0}.
\end{align}
Here, $n$ denotes the refractive index of the dye, and $c$ the speed of light \cite{Vretenar2021}.
Assuming a cavity composed of two planar mirrors, it immediately follows that a tilt between the mirrors, whereby $\Delta d(x,y)$ becomes a linear function, introduces a linear potential in the system.

The inverse problem arises when one considers the mode patterns as (experimentally) given and searches for the corresponding potential $V(x,y)$. Solving directly for $V(x,y)$, we obtain
\begin{equation} \label{eq:inverse-schroedinger}
\begin{aligned}
    V(x,y) &= -\frac{mc^2}{n^2} \frac{\Delta d(x,y)}{D_0}   \\
    &= E + \frac{\hbar^2}{2m} \frac{\nabla^2 \Psi(x,y)}{\Psi(x,y)}.
\end{aligned}
\end{equation}
\Cref{eq:inverse-schroedinger} is in general a two dimensional (2D) inverse problem.
For the intended application, however, which involves determining the cavity alignment, we can consider $\Delta d(x,y)$ and thus $V(x,y)$ to be additive, in the sense that 
\begin{align}
\Delta d(x,y) &=\Delta d_1(x)+\Delta d_2(y)\\
V(x,y) &= V_1(x) + V_2(y).
\end{align}
This allows to decompose the 2D problem into two 1D inverse problems
\begin{align}  
    \label{eq:pot_recon_1}
    V_1(x) &= -\frac{mc^2}{n^2} \frac{\Delta d_1(x)}{D_0}
    = E_1 + \frac{\hbar^2}{2m} \frac{\partial_x^2\Psi_1(x)}{\Psi_1(x)}\\
    \label{eq:pot_recon_2}
    V_2(y) &= -\frac{mc^2}{n^2} \frac{\Delta d_2(y)}{D_0}
    =E_2 + \frac{\hbar^2}{2m} \frac{\partial_y^2\Psi_2(y)}{\Psi_2(y)}
\end{align}
with $\Psi(x,y)=\Psi_1(x)\Psi_2(y)$ and $E=E_1+E_2$. Solving these two problems will determine the alignment of the microcavity along the $x-$axis, namely $\Delta d_1(x)$, and along the $y-$axis, namely $\Delta d_2(y)$, respectively.

\begin{figure*}[t]
    \includegraphics[width=0.475\textwidth]{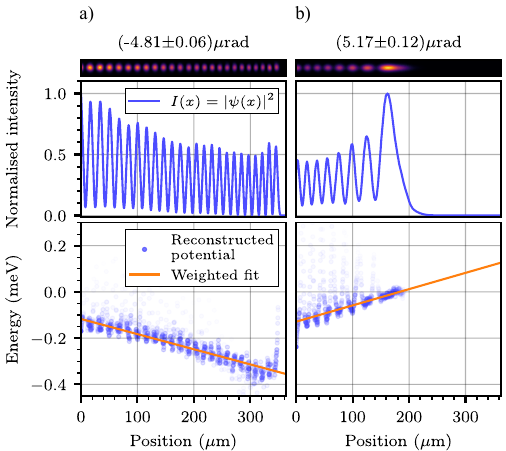}
    \hfill
    \includegraphics[width=0.475\textwidth]{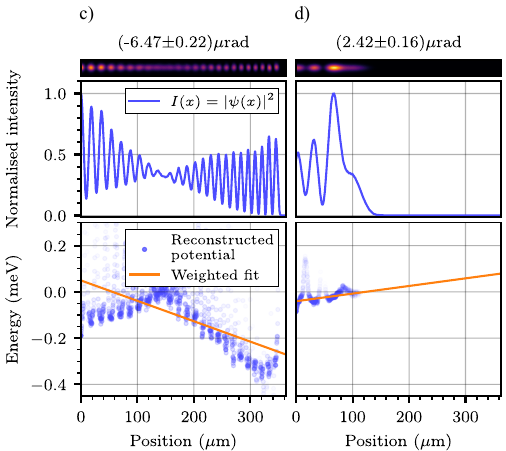}
    \caption{(a), (b), (c), (d) Mode patterns for different cavity alignments and/or pump spot positions. The top row shows four camera images of intensity patterns in the waveguide. The middle row shows the intensity integrated orthogonal to the waveguide axis (blue curve). The bottom row shows the reconstructed potential of that intensity distribution (blue circles) with a linear fit (orange line). The slope of the linear fit multiplied by the cavity length yields the cavity tilt (given on top with the error of the linear fit).}
    \label{fig:potential_reconstruction}
\end{figure*}

The resolution of this inverse problem demands careful consideration. At first glance, determining $V_1(x)$ using \cref{eq:pot_recon_1}, for instance, appears to necessitate knowledge of the full wavefunction, $\Psi_1(x)$. In contrast, experimental measurements typically yield $I_1(x) = |\Psi_1(x)|^2$. However, when considering modes that are confined within a finite volume, their wavefunction becomes real-valued, such that $\Psi_1(x) = s_1(x)\sqrt{I_1(x)}$ with a sign function $s_1(x)\in \{-1,+1\}$. Crucially, expressions of the form $\partial_{x,y}^2\Psi/\Psi$, as they appear in \cref{eq:pot_recon_1,eq:pot_recon_2}, are unaffected by the sign. Therefore, provided that the modes are spatially confined, knowing $|\Psi|$ is sufficient to solve the inverse problem. Another consideration is that expressions of the form $\partial_{x,y}^2\Psi/\Psi$ will become ill-defined at the nodes of the wavefunction, where $\Psi$ equals zero. It is crucial to exclude these points when determining the parameters of the unknown (here linear) potential function. Furthermore, it is necessary to ensure that the wavefunction employed to solve the inverse problem is dominated by a single eigenmode, not a superposition of eigenmodes possessing different energies. This will be investigated more closely in the following sections.

%%%%%%%%%%%%%%%%%%%%%%%%%%%%%%
%%%   EXPERIMENTAL SETUP   %%%
%%%%%%%%%%%%%%%%%%%%%%%%%%%%%%

\section{Experimental Setup}    \label{sec:experiment}

At the heart of our setup is an optically active microcavity comprised of two distributed Bragg reflectors (DBRs), one of which can be moved with three piezoelectric actuators with respect to the other, and an optical medium consisting of rhodamine 101 dissolved in ethylene glycol. One of the mirrors has an additional silicon (Si) layer between the DBR stack and the superpolished substrate, which is used to create almost arbitrary surface height profiles on the mirror using a direct laser writing technique \cite{vretenar2023}. A schematic of the setup can be seen in \Cref{fig:setup}(a) with a zoom-in on the microcavity in \Cref{fig:setup}(b). The cavity is non-resonantly pumped with a $\SI{532}{\nano m}$ pulsed laser.
A spatial light modulator (SLM) is used to move the pump spot in the plane of the cavity.
The light emitted from the cavity is split with a beam splitter and measured with a camera and a spectrometer.
The spectral signal is employed to track and stabilise the cavity length.

\Cref{fig:setup}(c) illustrates a top-view schematic of the nanostructured mirror, depicting two orthogonal linear waveguide structures terminated at both ends, named TX and TY. These surface structures are utilized to transversely confine light within an effectively 1D potential. This decomposes the 2D inverse problem not only mathematically but also physically into two 1D problems. The corresponding modes will be employed to ascertain the tilt on both axes using the inverse solution method introduced in Section \ref{sec:method}. \Cref{fig:setup}(d) presents the surface height profile and a cross section of such a structure. At one end of the waveguide structure, a ramp-like potential is introduced. By varying the pump spot position along this ramp, modes with different energies can be generated. For solving the inverse problem, we only use the signal from the flat part of the waveguides. \Cref{fig:setup}(e) shows examples of the measured intensity distributions, one for negative tilt and one for positive tilt. For positive tilts the intensity distributions do not necessarily extent over the full length of the waveguide and we effectively sample over a shorter distance.

%%%%%%%%%%%%%%%%%%%%%%%%%%%%%%%%
%%%   RESULTS & DISCUSSION   %%%
%%%%%%%%%%%%%%%%%%%%%%%%%%%%%%%%

\section{Results and Discussion}   \label{sec:results}

\begin{figure}[t!]
    \flushleft
    \includegraphics[width=0.45\textwidth]{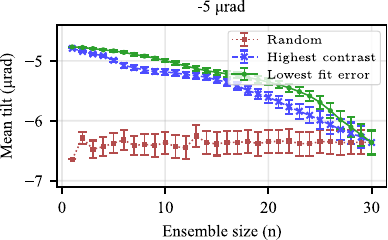}
    \caption{Average tilting angle as a function of ensemble size for a cavity alignment close to $\SI{-5}{\micro rad}$. By moving the pump spot along the ramp potential of the waveguide, we generate a set of 30 different (multi-)mode patterns. From these, we select $n$ patterns based on a certain selection criterion, solve the associated inverse problem, and calculate an average tilting angle. The selection process chooses the patterns according to specific criteria, namely the highest contrast (blue curve) and the lowest fit error (green curve). For comparison, these are juxtaposed with a random selection of modes (red curve). Each data point in this plot represents an average of 100 independent repetitions of this procedure. The error bar indicates the standard error.}
    \label{fig:tilt}
\end{figure}

The top row of \Cref{fig:potential_reconstruction} displays four examples of intensity distributions generated by the non-resonant optical pumping of the waveguide structures. 
The emission line is around a wavelength of $\SI{650}{\nano m}$. 
The middle row of \Cref{fig:potential_reconstruction} illustrates the intensities $I_{1,2}$ as a function of position. 
For this, the signal has been integrated orthogonal to the waveguide axis. 
We now solve the two 1D inverse Schrödinger equations in which we assume the potentials to be linear functions:
\begin{align}
     V_1(x) &=  a_1 x +  b_1 = E_1+ \frac{\hbar^2}{2m} \frac{\partial_x^2 \sqrt{I_1}}{\sqrt{I_1}}  \label{eq:pot_recon_x}   \\ 
     V_2(y) &=a_2 y +  b_2 = E_2+ \frac{\hbar^2}{2m} \frac{\partial_y^2 \sqrt{I_2}}{\sqrt{I_2}}.    \label{eq:pot_recon_y}
\end{align}
Note that for the tilt determination, we are only interested in the two slopes $a_{1,2}$. For the evaluation of the second derivatives, we smooth the data with a Savitzky–Golay filter and perform numerical differentiation. 
The results of this procedure are shown in the reconstructed potential energies (blue circles) in the bottom row of \Cref{fig:potential_reconstruction}. 
To obtain the slopes $a_{1,2}$, we apply a weighted linear fit on the data sets (orange lines), where high-intensity regions have higher weight, and lower intensity regions around the nodes of the wavefunction have lower weight. 
This is reflected in the shading of the blue circles.
The resulting slopes can be converted into the tilt angles shown at the top of the columns in \Cref{fig:potential_reconstruction}. 
The results presented in \Cref{fig:potential_reconstruction} indicate that our method can provide excellent estimates of the tilting angles, see, for example, \Cref{fig:potential_reconstruction}(a) and (b). 
Here, the method achieves an angle resolution as good as $\SI{60}{\nano rad}$ for the shown mode pattern. 
In fact, we consider this uncertainty to more likely reflect the non-flatness of our mirrors than a limitation of our method. 
The mirrors we use are based on superpolished substrates and have a roughness of $\SI{0.1}{\nano m}$ (root mean square).
Across the $\SI{360}{\micro m}$ length of our waveguides, a height variation of $\SI{0.1}{nm}$ corresponds to a tilting angle of $\SI{0.3}{\micro rad}$. Notably, this angle is larger than the reported angle resolution of $\SI{60}{\nano rad}$, demonstrating the method's high sensitivity.

In other cases, however, the tilt value has a somewhat larger uncertainty, or the procedure fails completely. We have identified mainly two mechanisms that can negatively affect the angle reconstruction. First, if the analyzed modes have a low spatial extension, then the reconstruction is obviously less accurate, see \Cref{fig:potential_reconstruction}(b) and (d). In our experiment, this mainly occurs at large positive angles. This does not pose a fundamental problem, however, as the modes would regain their full extension if the position of the optical pumping were moved to the opposite side of the waveguide.
The second mechanism is multimode excitation. The intensity distributions in \Cref{fig:potential_reconstruction}(c) and (d), for example, show the superposition of multiple eigenmodes with different energies, which violates one of the assumptions of our method.
It is clear that a reliable determination of the tilting angles needs to exclude such cases. To address this issue, we move the pump spot across different positions (typically around 30) using the SLM, scanning it through a defined path on the ramp to excite modes with different eigenenergies.
We then select intensity patterns that are closest to single mode emission, rather than a superposition of eigenmodes, according to a specific regularization criterion.
As the first criterion, we consider the contrast of the intensity distribution, that is, the normalized difference between local maxima and minima. If the contrast becomes small at some position, as in \Cref{fig:potential_reconstruction}(c) and (d), it indicates a superposition of multiple modes. The second criterion is the error encountered during the linear fitting of the reconstructed potential (derived from the covariance matrix).
We compare these criteria with a procedure in which we randomly select modes.

\begin{figure}[]
    \flushleft
    \includegraphics[width=0.45\textwidth]{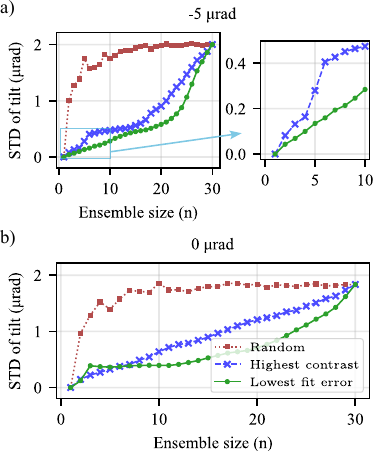}
    \caption{Standard deviation of the tilt angle measurement for different regularization schemes and two different cavity alignments, (a) $\SI{-5}{\micro rad}$ with a zoom-in on the $n\leq 10$ ensemble size, and (b) for $\SI{0}{\micro rad}$. The measurement procedure is the same as in \Cref{fig:tilt}. The tilt determination is found to work best for negative tilts, which create the modes with the largest spatial extension, see \Cref{fig:setup}(e).}
    \label{fig:tilt_STD}
\end{figure}

To test the different methods we prepare the microcavity in a specific tilt configuration using the piezoelectric actuators. As described before, we scan the pump spot across 30 different positions on the ramp and determine the tilt based on the excited intensity pattern.
In \Cref{fig:tilt}, we show the obtained tilt angles for a cavity alignment of $\SI{-5}{\micro rad}$ when averaging over the $n$ best modes according to the given selection criterion. Each data point shown in \Cref{fig:tilt} represents the average of 100 independent repetitions of this procedure, with error bars indicating the standard error of the mean. We observe a shift in the ensemble mean as the ensemble size $n$ increases and the selection process becomes less restrictive. 
In \Cref{fig:tilt_STD} we show the standard deviation of the measured tilting angle for two different cavity alignments. 
We find that for the random selection of modes, the standard deviation remains roughly constant, as expected. Conversely, the standard deviation is significantly lower when selecting modes with the highest contrast or lowest fitting error, indicating that these criteria lead to more consistent results in the reconstructed potential. Thus, selecting the best modes is found to be a useful procedure to enhance the determination of the cavity alignment.

%%%   SUMMARY   %%%

In summary, we introduce a new method for determining the alignment of a microcavity based on the inverse solution of the Schr\"odinger equation. The obtained angle uncertainty in our experiments is on the order of $\SI{100}{\nano\radian}$ per measurement cycle, where a single measurement cycle comprises 30 different intensity patterns created by 30 different pump spot positions. Selecting the intensity patterns for their single-mode nature is shown to be useful for achieving consistent measurement results. We would like to emphasize that our method is broadly applicable to many types of optical resonators, where the specific details of the implementation can significantly differ from the one demonstrated here. For example, it is neither necessary for the resonator to be pumped non-resonantly, nor for the pump position to be changeable. A minimal implementation of our method could simply involve generating a transversally confined mode through resonant excitation and analyzing it in the manner presented here. We expect that a main application area of our method will be research on two-dimensional photonic or polaritonic gases in microresonators \cite{Kalinin2022,Inagaki2016,Berloff2017,Alyatkin2020,Stroev2020,Stroev2021}. Such systems can, for example, be used as analog simulators for spin systems \cite{Verstraelen2024, Hur2024, Stroev2023, Pierangeli2019, Kalinin2020}. In this context, the homogeneity of the spin network and, thus, the alignment of resonators play an essential role. Another area of application could be high-precision interferometry, such as gravitational wave sensing, where a detailed understanding of the alignment of the optical components is crucial \cite{Acernese_2015,Aso2013,Hartig2023,Weaver2022}.

%%%%%%%%%%%%%%%%%%%%%%%%%%%%
%%%   ACKNOWLODGEMENTS   %%%
%%%%%%%%%%%%%%%%%%%%%%%%%%%%

\begin{acknowledgments}

This work has received funding from the European Research Council (ERC) under the European Union’s Horizon 2020 research and innovation program (Grant Agreement No. 101001512) and from the NWO (grant no. OCENW.KLEIN.453).

\end{acknowledgments}

%%%%%%%%%%%%%%%%%%%%
%%%   APPENDIX   %%%
%%%%%%%%%%%%%%%%%%%%

%\appendix

%\section{Appendixes}
%If there is only one appendix, then the letter ``A'' should not
%appear. This is suppressed by using the star version of the appendix
%command (\verb+\appendix*+ in the place of \verb+\appendix+).

%\section{A little more on appendixes}

%\subsection{\label{app:subsec}A subsection in an appendix}

%\nocite{*}

\bibliographystyle{apsrev4-1}
\bibliography{bibliography} % Produces the bibliography via BibTeX.

\end{document}